\documentclass[a4paper,fleqn]{cas-sc}

\usepackage[numbers]{natbib}
\usepackage[section]{placeins}
\def\tsc#1{\csdef{#1}{\textsc{\lowercase{#1}}\xspace}}
\tsc{WGM}
\tsc{QE}
\tsc{EP}
\tsc{PMS}
\tsc{BEC}
\tsc{DE}
\begin{document}
\let\WriteBookmarks\relax
% Official CAS float parameters (from elsdoc-cas.tex)
\setcounter{topnumber}{2}
\setcounter{bottomnumber}{2}
\setcounter{totalnumber}{4}
\renewcommand{\topfraction}{0.85}
\renewcommand{\bottomfraction}{0.85}
\renewcommand{\textfraction}{0.15}
\renewcommand{\floatpagefraction}{0.7}

% Short title
\shorttitle{Scaling Flexibility using Dynamic Tariffs}

% Short author
\shortauthors{L. Brylle et~al.}

% Main title of the paper
\title [mode = title]{Scaling Demand-Side Flexibility Through Dynamic Tariffs}                      
% Title footnote mark
% eg: \tnotemark[1]
%\tnotemark[1,2]

% Title footnote 1.
% eg: \tnotetext[1]{Title footnote text}
% \tnotetext[<tnote number>]{<tnote text>} 
%\tnotetext[1]{This document is the results of the research
%   project funded by the National Science Foundation.}

%\tnotetext[2]{The second title footnote which is a longer text matter
%   to fill through the whole text width and overflow into
%   another line in the footnotes area of the first page.}

% First author
%
% Options: Use if required
% eg: \author[1,3]{Author Name}[type=editor,
%       style=chinese,
%       auid=000,
%       bioid=1,
%       prefix=Sir,
%       orcid=0000-0000-0000-0000,
%       facebook=<facebook id>,
%       twitter=<twitter id>,
%       linkedin=<linkedin id>,
%       gplus=<gplus id>]
\author[1]{Lucas Brylle}[%type=editor,
%                       auid=000,
%                        prefix=Sir,
%                        role=Researcher,
                        orcid=0009-0004-2740-9217]

% Corresponding author indication
%\cormark[1]

% Footnote of the first author
%\fnmark[1]

% Email id of the first author
\ead{lucasbrylle6@gmail.com}

% URL of the first author
%\ead[url]{www.cvr.cc, cvr@sayahna.org}

%  Credit authorship
\credit{Conceptualization, Writing - original draft, Methodology, Visualization, Software}

% Address/affiliation
\affiliation[1]{organization={Department of Applied Mathematics and Computer Science},
    addressline={Technical University of Denmark (DTU), Richard Pedersens Plads, Building 324}, 
    city={2800 Lyngby},
    % citysep={}, % Uncomment if no comma needed between city and postcode
 %   postcode={}, 
    % state={},
    country={Denmark}}

% Second author
\author[2]{Niels Andersen}[]

\credit{Data curation, Supervision, Resources, Conceptualization}

% Third author
\author[1]{Henrik Madsen}[orcid=0000-0003-0690-3713
   ]
%\fnmark[2]
\cormark[1]
\ead{hmad@dtu.dk}
\ead[URL]{www.henrikmadsen.org}

\credit{Methodology, Supervision, Writing - review and editing}

% Address/affiliation
\affiliation[2]{organization={Cerius-Radius},
    addressline={Teknikerbyen 25}, 
%    postcode={2830},
    city={2830 Virum},
     %citysep={}, % Uncomment if no comma needed between city and postcode
%    postcode={2830}, 
%    state={Trivandrum},
    country={Denmark}}

% Fourth author
%\author%
%[1,3]
%{Henrik Madsen}
%\cormark[2]
%\fnmark[1,3]
%\ead{rishi@stmdocs.in}
%\ead[URL]{www.stmdocs.in}

%\affiliation[3]{organization={STM Document %Engineering Pvt Ltd.},
%    addressline={Mepukada}, 
%    city={Malayinkil},
%    % citysep={}, % Uncomment if no comma %needed between city and postcode
%    postcode={695571}, 
%    state={Trivandrum},
%    country={India}}

% Corresponding author text
\cortext[cor1]{Corresponding author}
%\cortext[cor2]{Principal corresponding author}

% Footnote text
%\fntext[fn1]{This is the first author footnote. but is common to third
%  author as well.}
%\fntext[fn2]{Another author footnote, this is a very long footnote and
%  it should be a really long footnote. But this footnote is not yet
%  sufficiently long enough to make two lines of footnote text.}

% For a title note without a number/mark
%\nonumnote{This note has no numbers. In this work we demonstrate $a_b$
%  the formation Y\_1 of a new type of polariton on the interface
%  between a cuprous oxide slab and a polystyrene micro-sphere placed
%  on the slab.
%  }

% Here goes the abstract
\begin{abstract}
\noindent The ongoing electrification and integration of renewable energy sources in Denmark's distribution grids pose significant operational challenges, including insufficient reserve capacity, component degradation due to overload, voltage instability, and increasing infrastructure investment requirements. This article argues that implicit demand-side flexibility (DSF) incentivized through dynamic tariffs offers the most scalable and cost-effective approach to address these challenges in a modern distribution network. We demonstrate that while explicit flexibility mechanisms provide operational certainty, they cannot scale to address system-wide congestion across heterogeneous customer bases. Drawing on empirical consumption data showing strong price-responsive behavior, varying prices due to, e.g., regulatory frameworks including the Danish Market Model 3.0 and Tariff Model 3.0, and economic analysis, we demonstrate potential grid savings of 13--48 million DKK per constrained substation through deferred or avoided reinforcement. We argue that implicit DSF mechanisms represent the necessary pathway for revenue-neutral scalable flexibility solutions that can defer costly grid reinforcements while maintaining system reliability. Beyond direct grid savings, additional value streams include avoided peak generation costs, reduced connection delays, and lower outage risk, further strengthening the economic case. Critically, dynamic tariffs offer the mechanism through which real-time grid constraints can be communicated to consumers, enabling price signals that accurately reflect the actual state of the capacity of the distribution network at any given point in time and space.

\end{abstract}

% Use if graphical abstract is present
% \begin{graphicalabstract}
% \includegraphics{figs/grabs.pdf}
% \end{graphicalabstract}

% Research highlights
\begin{highlights}
\item Value of flexibility relative to traditional grid reinforcement
\item Study of implicit versus explicit flexibility
\item Dynamics tariffs for unlocking flexibility at scale
\item Cost-reflective grid tariffs
\item Europe's CAPEX-driven  regulatory approach systematically favors infrastructure expansion over smarter digital solutions 
\end{highlights}

% Keywords
% Each keyword is seperated by \sep
\begin{keywords}
demand-side flexibility \sep distribution grids \sep 
dynamic tariffs \sep deferred grid costs \sep scalable flexibility
\end{keywords}

\maketitle

\section{Introduction}
\label{sec:intro}

The Danish electricity grid faces a fundamental mismatch between the pace of electrification and the timeline for traditional infrastructure expansion. Accelerating adoption of electric vehicles, heat pumps, data-centers, distributed generation, and behind-the-meter battery storage has introduced operational challenges including insufficient reserve capacity, component degradation from sustained overloading, voltage instability, and escalating investment requirements. Traditional responses, such as grid reinforcement through new cables and transformers, are capital-intensive and time-consuming, creating a growing gap between grid capacity and demand growth.

This gap cannot be closed through traditional means alone. The Danish distribution grid investment needs are expected to rise from approximately DKK 2 billion to DKK 6.7 billion annually 
over the coming decade \cite{GreenPowerDenmark2024}. Similarly, 
the costs of grid balancing services have escalated sharply: before 
2020, these costs were stable at 600--900 million DKK annually, but 
Energinet projects the demand for such services to increase by 150\% 
toward 2040 \cite{Energinet2024Outlook}. Denmark is experiencing a dramatic increase in hours with negative electricity prices. From January to May 2025, Danish solar parks faced negative prices during 18\% of operating hours, up from 13\% during the same period in 2024 \cite{ingenioeren2025negativeprices}. At this scale, alternative approaches are not merely desirable but necessary. According to DNV's EU-wide analysis of demand-side flexibility potential, the traditional infrastructure-centric approach will be 
too costly compared with digitally enabled demand-side flexibility \cite{DNV2022FlexibilityEU}.

The consequences of relying on construction alone are already visible in the connection queue. Energinet has reported waiting times of up to ten years for new grid connections, with several hundred industrial and commercial projects queued for capacity that does not yet exist \cite{Steno2026GreenPowerQueue}. The Danish National Audit Office has documented that roughly 70\% of Energinet's grid-expansion projects are delayed, that delayed projects take on average around 80\% longer than planned, and that the build-out rate must more than double, from roughly 120 to 250 km of new high-voltage line per year, if national 2050 targets are to be met \cite{Rigsrevisionen2026Energinet}. This widening gap between connection demand and physical delivery makes it untenable to treat infrastructure expansion as the primary lever; flexibility must absorb load growth in the years before reinforcement can arrive.

These challenges have led to a broader international trend toward ``flexibility-first'' grid management. In the United Kingdom, the Office of Gas and Electricity Markets (Ofgem) has implemented the RIIO (Revenue = Incentives + Innovation + Outputs) framework, which regulates distribution and transmission network operators through a performance-based model designed to promote smarter, more innovative networks \cite{OfgemRIIOHandbook2010}. Under the RIIO-ED2 price control (2023--2028), Ofgem introduced a dedicated DSO incentive that further rewards DSO for efficiently developing their networks using flexible alternatives to traditional reinforcement, with financial rewards tied to flexibility procurement and stakeholder satisfaction \cite{OfgemRIIOED2FD2022}. In California, the California Independent System Operator (CAISO) has documented growing operational challenges from high solar penetration through the widely cited ``duck curve,'' where midday oversupply from distributed photovoltaics creates steep evening ramping requirements of up to 13,000 MW in three hours \cite{CAISO2016DuckCurve}. These grid conditions have prompted the California Public Utilities Commission (CPUC) to mandate time-of-use rate reform and to investigate real-time pricing mechanisms that can signal local distribution constraints down to the substation level \cite{CPUC2022CalFUSE}. In South Australia, where more than one-third of homes host distributed solar PV systems, SA Power Networks has rolled out dynamic operating envelopes and flexible export connections as standard options since July 2023, using real-time network signals to manage local voltage and thermal constraints from high DER penetration \cite{SAPNFlexibleExports2024}. By situating the Danish case within this global context, we demonstrate that implicit flexibility via dynamic tariffs is not merely a local experiment but a core component of a global transition toward smarter, more cost-efficient distribution grids.

It is important to distinguish between static time-of-use (ToU) tariffs and truly dynamic tariffs. Current ToU structures in Denmark apply predetermined price differentials based on typical daily consumption patterns, meaning higher prices during expected peak hours and lower prices overnight. While ToU tariffs represent an improvement over flat rates, they cannot address the fundamental challenge: \textit{grid constraints are location-specific and time-varying in ways that static schedules cannot capture}. A transformer may be constrained on a cold January evening when many heat pumps operate simultaneously, but unconstrained on a mild autumn day at the same hour. Dynamic tariffs that respond to actual grid conditions in real-time rather than predetermined schedules, are necessary to align customer behavior with the true state of distribution network capacity.

Demand-side flexibility (DSF) offers a complementary approach by encouraging consumers to adjust electricity usage in response to grid conditions, potentially deferring costly expansions while enhancing system resilience. However, DSF can be mobilized through fundamentally different mechanisms, explicit and implicit, with dramatically different scalability characteristics.

This article argues that implicit DSF through dynamic tariffs is a promising and viable way to achieve system-wide flexibility at the scale required to address Denmark's distribution grid challenges. We support this argument through four propositions:

\begin{enumerate}
    \item Explicit flexibility mechanisms, while valuable for specific applications, cannot scale to address widespread, moderate congestion across heterogeneous customer bases.
    \item Empirical evidence demonstrates that customers already exhibit strong price-responsive behavior under existing wholesale price exposure. Consequently, the behavioral foundation for implicit flexibility exists.
    \item Danish regulatory frameworks (Market Model 3.0 and Tariff Model 3.0) have established essential preconditions: universal hourly settlement, standardized time-differentiated tariffs, capacity-based charging, and independent aggregator access, but the current implementation remains limited to static time-of-use structures. The step from predetermined ToU schedules to truly dynamic tariffs that reflect real-time and location-specific grid constraints represents the critical remaining gap \cite{DanishTariffModel3,DanishMarketModel3}.
    \item The proliferation of automated, price-responsive devices (EVs, heat pumps, smart appliances) will transform implicit flexibility from probabilistic to reliable as automation becomes the norm rather than the exception.
\end{enumerate}

The analysis draws on empirical consumption data from the Cerius-Radius network, Denmark's largest distribution network serving 1.5 mill consumers in the Eastern part of Denmark, including the Greater Copenhagen area, to demonstrate these propositions and quantify the economic value of implicit flexibility relative to traditional reinforcement.

\section{Why Explicit Flexibility is no longer sufficient}

Demand-side flexibility in distribution grid management refers to the modification of consumer electricity demand in response to signals from the grid operator. This flexibility can be mobilized through two fundamentally different approaches: explicit and implicit mechanisms. Understanding their structural differences reveals why implicit mechanisms are necessary, not merely preferable.

\subsection{Explicit Flexibility: Reliable but Inherently Limited}
Explicit flexibility involves direct contractual arrangements in which customers commit to providing specific demand reduction services in exchange for compensation. Distribution System Operators (DSOs) can dispatch these resources with defined reliability, as participating customers agree to curtailment rules during grid stress periods. Compensation typically includes reduced connection fees, capacity payments, or direct remuneration for activated flexibility.

Denmark's DSOs have developed several explicit mechanisms:

\subsubsection*{Limited Grid Access (Begrænset Netadgang - BNA)} offers customers connecting at 10 kV or higher voltage levels reduced connection fees and expedited grid access in exchange for accepting curtailment obligations. However, BNA eligibility requires that sufficient capacity exists for the customer's baseline consumption. It enables faster connection of new loads but can not defer reinforcement where capacity is already exhausted. BNA operates in the buffer space of existing capacity; it is not a capacity relief mechanism \cite{RadiusBNA,RadiusLimitedGridAccessAgreement2019}.

\subsubsection*{Market-Based Activation of Flexibility (MAF)} provides contracted downward regulation within geographically defined areas, procured through competitive tender. While MAF offers greater operational discretion than BNA, it requires bilateral procurement, verified delivery against baselines, and compensation structures that increase transaction costs \cite{GreenPowerDenmark2024LocalTariff}.

The fundamental limitation of explicit mechanisms is structural: they require bilateral contract negotiation, active customer engagement, facilities capable of meaningful load reduction, and ongoing verification. These requirements inherently limit participation to customers with:
\begin{itemize}
    \item Sufficient load magnitude to justify transaction costs
    \item Technical capability for controllable reduction
    \item Organizational capacity for contractual commitments
    \item Willingness to accept dispatch obligations
\end{itemize}

Furthermore, many potential flexibility providers have primary operational objectives that supersede energy cost optimization. For example, wastewater treatment plants, despite their substantial and often flexible electricity loads, do prioritize their core function of processing wastewater over participating in flexibility schemes \cite{brok2020a,stentoft2021a}. Data centers are in the same category. This operational reality means that even facilities with significant technical flexibility potential are seldom willing to commit to explicit flexibility arrangements that could interfere with their primary mission.

For DSOs facing congestion across residential areas with distributed flexible loads, which represent the dominant use case in electrifying distribution grids, explicit contracting is impractical at scale.

\subsection{Implicit Flexibility: The Scalable Alternative}

Implicit flexibility relies on price signals to encourage voluntary behavioral changes without direct operator control or dispatch. Once a price signal structure is established, participation requires no additional contracts, bilateral negotiations, or control systems. The response emerges from the aggregate behavior of the price-responsive customers \cite{corradi2013a}.

\textbf{Scalability is the defining advantage of implicit mechanisms.} Unlike explicit approaches that require per-customer transaction costs, implicit flexibility scales with zero marginal administrative burden per additional participant. A dynamic tariff structure that serves 10,000 customers can serve 100,000 or 1,000,000 with the same infrastructure. This scalability derives from the fundamental design: rather than negotiating individual curtailment agreements, the DSO broadcasts a price signal and allows customers to self-select their response based on their individual circumstances and preferences.

Table~\ref{tab:comparison} summarizes the key distinctions. The fundamental trade-off between reliability and scalability defines where each mechanism is appropriate. Explicit mechanisms suit large customers with significant, controllable loads; implicit mechanisms address the long tail of distributed flexibility that explicit approaches cannot reach.

\begin{table}[htbp]
\centering
\caption{Comparison of Implicit and Explicit Demand-Side Flexibility Approaches}
\label{tab:comparison}
\begin{tabular}{@{}lll@{}}
\toprule
Aspect & Implicit (Price-Driven) & Explicit (Contracted) \\
\midrule
Activation mechanism & Price signals & Direct dispatch \\
Scalability & System-wide & Limited to participants \\
Administrative burden & Low (once established) & High (per customer) \\
Customer commitment & Voluntary response & Binding agreement \\
Appropriate for & Distributed loads & Large, controllable loads \\
Addresses existing congestion & Yes & Limited (BNA: No) \\
\bottomrule
\end{tabular}
\end{table}

\subsection{The Local Collective Tariff: A Hybrid Approach}

The local collective tariff scheme constitutes an implicit DSF product designed to incentivize coordinated consumption and production behavior within geographically bounded groups of low-voltage customers \cite{GreenPowerDenmark2024LocalTariff}. Eligible customers form a local collective under a shared virtual metering point, where network tariffs are calculated on netted consumption and production.

The tariff structure combines a capacity-based power tariff accounting for approximately 75\% of network cost recovery, with a reduced time-differentiated energy tariff. This design shifts cost exposure from volumetric consumption toward coincident peak demand, thereby incentivizing collectives to reduce maximum load and improve synchronization between local generation and consumption. The high capacity share reflects the primary objective of managing local capacity constraints, though it correspondingly reduces the marginal price signal for intra-day load shifting.

\begin{figure}[pos=!htb]
    \centering
    \includegraphics[width=0.85\linewidth]{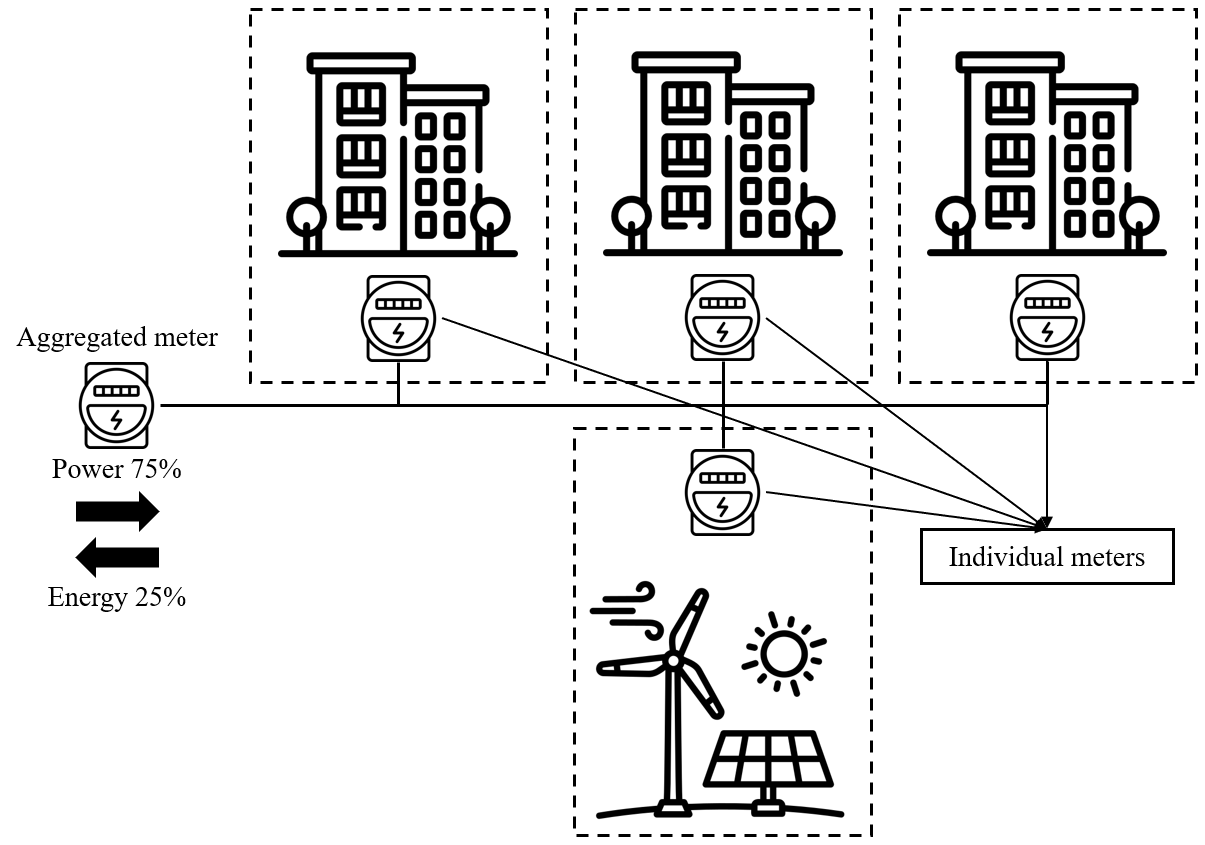}
    \caption{An overview of The Local Collective Tariff Framework}
    \label{fig:local_collective}
\end{figure}

While promising, collective tariffs require customer coordination and self-selection, limiting their reach compared to system-wide dynamic tariffs, but offers a great stepping stone to a more implicit future.

\section{Technical Evidence: Price Response Already Exists}

The viability of implicit flexibility depends on whether customers actually respond to price signals. Empirical evidence from the Cerius/Radius network demonstrates that substantial price-responsive behavior already exists under current wholesale price exposure, even without DSO-specific dynamic tariffs.

\subsection{Industrial Facilities: Strong Price Correlation}

\begin{figure}[pos=!htb]
    \centering
    \includegraphics[width=0.95\linewidth]{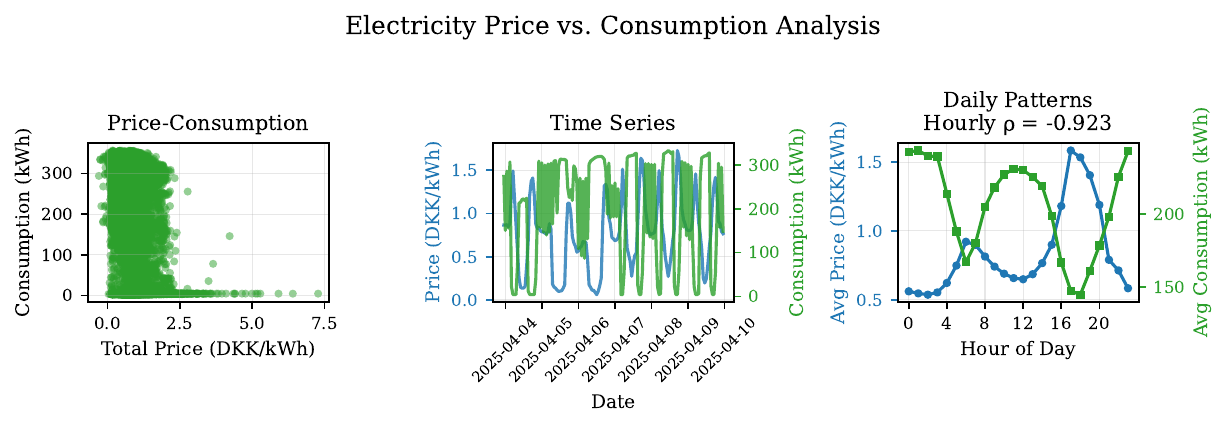}
    \caption{Price-consumption analysis for an industrial facility. Left panel shows hourly scatter plot of consumption versus spot price; center panel shows time series of hourly spot price and consumption; right panel shows average daily patterns. The correlation coefficient ($\rho = -0.923$) is computed from the hourly price-consumption relationship shown in the left panel. Note: Prices shown are Nord Pool spot prices plus network tariffs; the total price faced by consumers would additionally include taxes and retailer margins.}
    \label{fig:industrial}
\end{figure}

Figure~\ref{fig:industrial} presents consumption data from a major industrial facility connected to a secondary substation. The facility demonstrates strong inverse correlation between hourly electricity prices and consumption ($\rho = -0.923$), with consumption decreasing significantly during high-price periods and sub-processes halting entirely during price spikes.

This voluntary adaptation occurs without contractual obligations or direct operator dispatch, illustrating how dynamic tariffs can foster scalable flexibility among industrial users through market-based incentives rather than explicit controls.

\subsection{Residential Sector: The EV Challenge as Opportunity}

\begin{figure}[pos=!htb]
    \centering
    \includegraphics[width=0.90\linewidth]{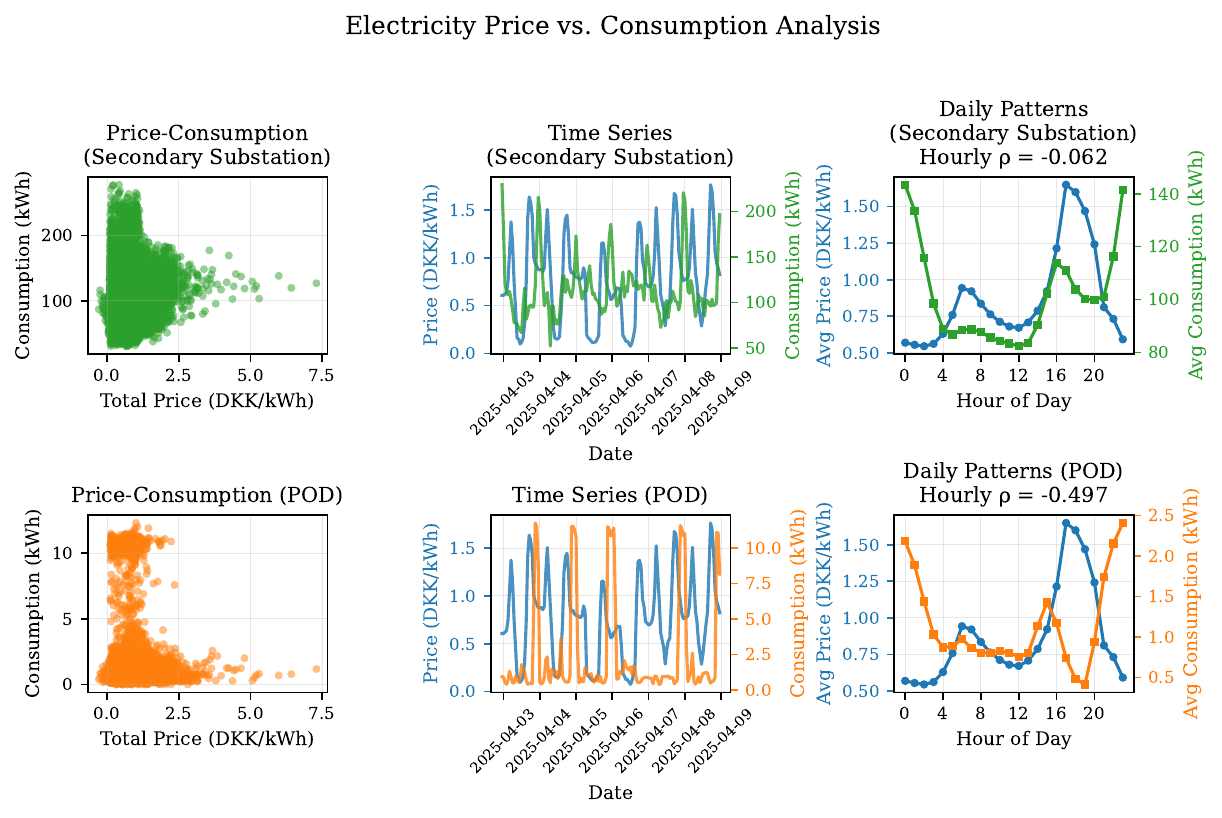}
    \caption{Price-consumption analysis for a residential area with high EV penetration. Top row: aggregated consumption at the secondary substation level. Bottom row: consumption at an individual point of delivery (POD), representing a single EV charging point. Prices shown are Nord Pool spot prices plus network tariffs (currently static ToU); actual consumer-facing prices would additionally include taxes and retailer margins. The strong price response observed under current pricing suggests even greater response potential when DSO-level dynamic tariffs replace the existing static tariff component. Charging concentrated during low-price hours causes daily capacity exceedance up to 50\% above rated transformer limits ($\rho = -0.497$ for individual charging point).}
    \label{fig:residential}
\end{figure}

Price-responsive behavior in the residential sector reveals both the potential and the necessity of DSO-level dynamic tariffs. Figure~\ref{fig:residential} shows consumption at a secondary substation in a residential area with high electric vehicle penetration. Charging activities align closely with low wholesale prices, resulting in daily capacity exceedance of up to 50\% above rated transformer limits.

This pattern reveals a critical insight: \textbf{customers are already price-responsive, but the current tariff component is static and does not reflect the actual and local grid constraints}. When combined prices are low, customers rationally shift consumption to these periods. However, because the network tariff follows a fixed time-of-use schedule rather than actual grid conditions, this rational response can create local grid stress when transformer capacity is already constrained.

The solution is not to suppress the price response but to \textit{replace the static tariff component with dynamic signals that reflect distribution grid  constraints}. DSO dynamic tariffs should complement wholesale price exposure by incorporating local (in space and time) congestion costs, thereby aligning price signals with grid reality at all voltage levels. Such methods are described in \cite{almassalkhi2026a}. The same price-responsive behavior that currently causes local overloading would instead contribute to local congestion relief \cite{Madsen2024FlexibilityIncentivising}. The EV charging pattern in Figure~\ref{fig:residential} demonstrates that implicit flexibility \textit{can} work and that the behavioral response to prices already exists.

\begin{figure}[pos=!htb]
    \centering
    \includegraphics[width=0.90\linewidth]{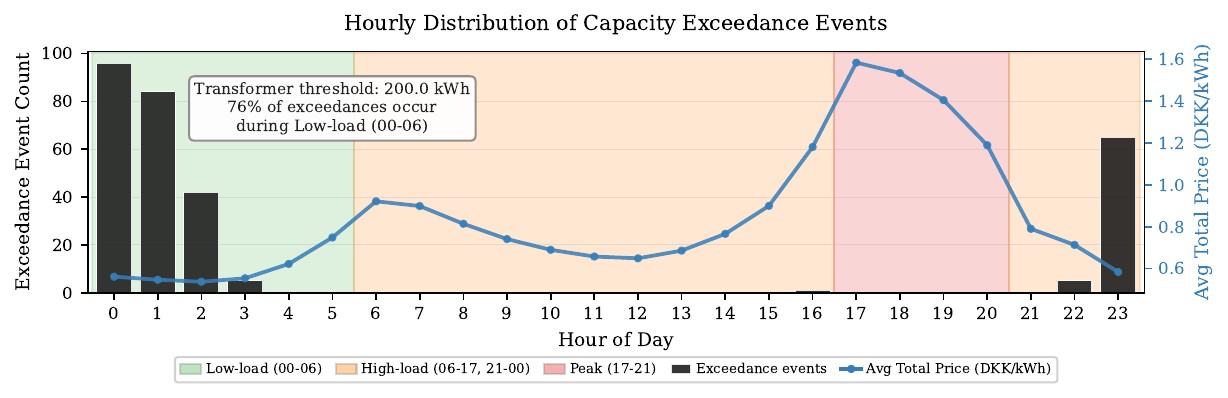}
    \caption{Hourly distribution of capacity exceedance events at a residential secondary substation with high EV penetration, with average total electricity price (Nord Pool spot price plus network tariff) overlaid. The transformer threshold is 200~kWh. Shaded regions indicate the current ToU tariff periods: low-load (00:00--06:00), high-load (06:00--17:00 and 21:00--00:00), and peak (17:00--21:00). Of all exceedance events, 76\% occur during the low-load period, when prices are lowest. This illustrates the fundamental misalignment between static ToU tariffs and actual local grid constraints.}
    \label{fig:congestion_timing}
\end{figure}

The timing of congestion events further underscores the inadequacy of static tariff structures. Figure~\ref{fig:congestion_timing} presents the hourly distribution of capacity exceedance events at the residential substation, overlaid with the average total price (spot plus network tariff) by hour of day. The mismatch is striking: 76\% of all exceedance events occur during the low-load period (00:00--06:00), precisely when prices are at their daily minimum. Under the current ToU tariff schedule, these hours are classified as ``off-peak,'' meaning the tariff component actively \textit{encourages} the concentrated EV charging that causes local grid stress. Meanwhile, the peak tariff period (17:00--21:00), when prices are highest, sees almost no congestion at this substation.

This result powerfully illustrates why static ToU tariffs are insufficient for managing distribution-level congestion. Dynamic tariffs that reflect real-time local constraints would instead signal high prices precisely when transformers approach capacity limits, regardless of the hour redirecting the same price-responsive behavior from a source of grid stress into a tool for congestion relief.

\subsection{Elasticity Distribution: Industrial Customers as Pilot Candidates}

\begin{figure}[pos=!htb]
    \centering
    \includegraphics[width=0.80\linewidth]{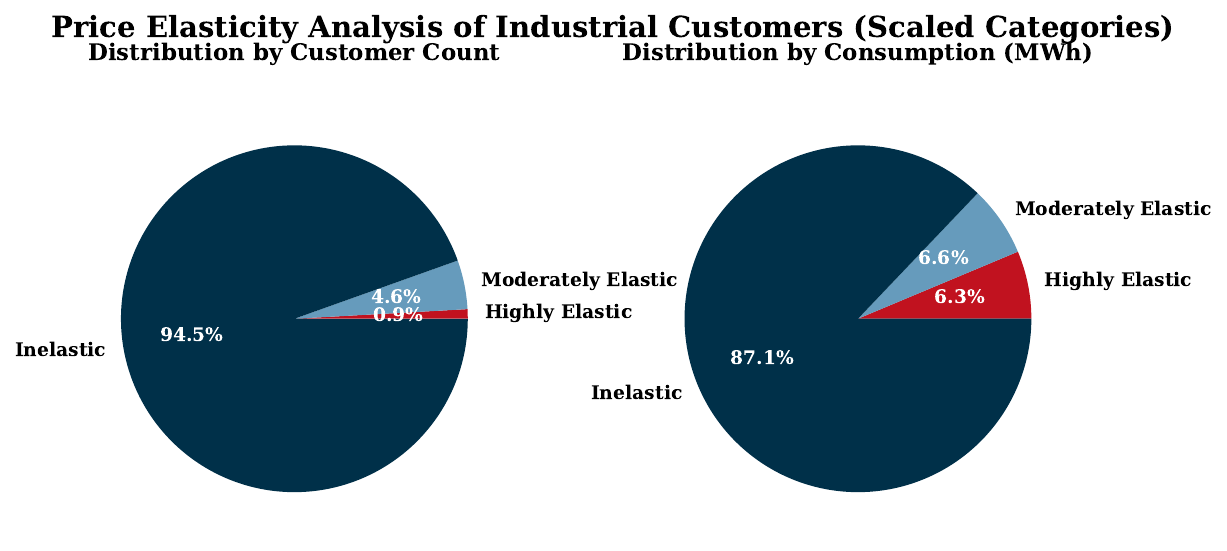}
    \caption{Price elasticity distribution among industrial customers in the Cerius-Radius network. Left: distribution by customer count. Right: distribution by consumption volume (MWh). Data source: Analysis of hourly metering data from industrial customers connected to the Cerius-Radius distribution network, 2023--2025.}
    \label{fig:elasticity}
\end{figure}

The analysis presented in this section focuses specifically on industrial customers in the Cerius-Radius network. This segment is particularly relevant for initial pilot programs because industrial facilities typically have (i) metering infrastructure capable of capturing price-responsive behavior, (ii) controllable processes that enable measurable load shifting, and (iii) commercial incentives that motivate active participation. Figure~\ref{fig:elasticity} presents the distribution of price responsiveness across this industrial customer base, 
noting that these distributions reflect current automation levels (2023--2025) and are likely conservative as adoption of automated price-responsive technologies increases.

While 94.5\% of industrial customers exhibit inelastic demand, the 5.5\% with elastic demand account for 12.9\% of total consumption. Highly elastic customers consume disproportionately more energy relative to their numbers (6.3\% of consumption vs. 0.9\% of customers), indicating that large industrial loads are more likely to have operational flexibility enabling price response.  

This concentration has important implications for the viability of implicit flexibility and on how we implement a stepping stone towards fully dynamic tariffs:

\begin{enumerate}
    \item \textbf{Flexibility potential exceeds typical congestion requirements}: Distribution grid congestion events are typically concentrated in a small fraction of annual hours. In the Cerius-Radius network, constrained substations require demand reduction during approximately 3--4\% of annual hours at most. The 12.9\% of industrial consumption exhibiting price-responsive behavior represents hundreds of MW of potential load flexibility, which substantially exceeds these requirements.
    
    \item \textbf{Implicit mechanisms automatically concentrate response where flexibility exists}: Unlike explicit mechanisms requiring DSO identification and recruitment of flexible customers, dynamic tariffs implies that customers are in the center; they can always decide.
    
    \item \textbf{The ``inelastic majority'' is not a barrier}: System-wide flexibility does not require universal participation. Even if most customers do not actively adjust consumption, the aggregate response from price-elastic segments can be sufficient for grid management.
\end{enumerate}

A key advantage of implicit flexibility through dynamic tariffs is that customers remain in control of their consumption decisions at all times. Unlike explicit mechanisms that involve dispatch obligations, dynamic tariffs simply provide price signals that reflect grid conditions, customers can always decide whether and how to respond based on their individual circumstances and preferences. This customer-centric design is fundamental: dynamic tariffs do not compel behavior change but rather create incentives that allow customers who \textit{can} respond to benefit from doing so, while those who cannot or choose not to simply pay the prevailing tariff.

This principle also informs the pathway toward fully dynamic tariffs. The evidence presented in this section suggests that industrial customers with highly elastic demand represent ideal candidates for initial pilot programs. These pilots can validate tariff design, demonstrate grid benefits, and build regulatory confidence before broader deployment. The local collective tariff (Section~2.3) offers another stepping stone, allowing geographically bounded groups to experience capacity-based pricing before system-wide implementation. Each step provides learning opportunities for DSOs, aggregators, and regulators on how best to design and implement increasingly dynamic tariff structures.

\section{Regulatory Frameworks Enabling DSF}

The deployment of demand-side flexibility ultimately depends on the regulatory environment that shapes DSO incentives, obligations, and operational freedom. Denmark has established a regulatory framework that explicitly enables implicit flexibility deployment.

\subsection{European Union Framework}

The legal foundation for DSF in Europe is established in the Clean Energy for All Europeans package, specifically Directive (EU) 2019/944 and Regulation (EU) 2019/943 \cite{EU2019_944,EU2019_943}. Article 32 of Directive 2019/944 introduces a pivotal requirement: DSOs must consider flexibility services, including demand response, storage, and distributed generation, as potential alternatives to network expansion when developing grid investment plans. This ``flexibility-first'' principle requires DSOs to evaluate non-wire alternatives before proceeding with traditional reinforcement.

Regulation 2019/943 complements this by mandating that congestion management and re-dispatching follow market-based, non-discriminatory principles. Dynamic pricing contracts become a formal consumer right under Directive 2019/944, ensuring customers with smart meters can access supply products exposing them to time-varying prices.

The 2024 Electricity Market Design reform, comprising Directive (EU) 2024/1711 and Regulation (EU) 2024/1747, strengthens these provisions while improving consumer protections \cite{EU2024_1711,EU2024_1747}.

\subsection{Danish Implementation}

Denmark has adopted a proactive approach to DSF, implementing several EU provisions ahead of mandatory timelines. A critical enabler was the nationwide deployment of smart meters and universal hourly settlement, completed by 2020 \cite{SmartMeterRolloutDK}. This infrastructure provides the data granularity essential for cost-reflective tariffs and automated demand response.

Central to Danish regulation is \textbf{Tariff Model 3.0}, developed by the electricity sector and approved by Forsyningstilsynet (the Danish Utility Regulator). The model explicitly supports cost-reflective, time-differentiated, and capacity-based tariffs while preserving revenue neutrality for DSOs across regulatory periods \cite{DanishTariffModel3}. Subsequent regulatory decisions have approved DSO-led time-of-use and capacity components \cite{DanishTariffModel3}, enabling tariff signals for congestion management consistent with EU requirements.

\textbf{Market Model 3.0} complements these tariff reforms by clarifying roles and data flows between DSOs, retailers, and aggregators, facilitating the integration of flexibility services into routine grid operations. The main change relaxes the previous mandate that required each aggregator to be associated with a retailer and a balance responsible party, reducing barriers to aggregator market participation. This implies that multiple aggregators can now provide services for a single household, though in practice, integrated Home Energy Management Systems (HEMS) that coordinate all assets behind a single meter are likely to deliver more effective optimization than independent aggregators operating on individual devices.

The emergence of consumer-facing flexibility products demonstrates that market actors are already positioning for this future. Andel Energi, Denmark's largest energy company, launched a home battery product in late 2025 that enables customers to optimize consumption against time-varying prices and participate in grid balancing services through Energinet \cite{AndelEnergiBattery2025}. Customers with batteries can earn revenue by allowing their storage to be dispatched during system stress periods, with Andel handling the market interface. Although this represents explicit flexibility at the device level, the customer-facing proposition is implicit: the battery responds automatically to price signals without requiring active decision-making from the homeowner. Such hybrid products explicit in their technical operation but implicit in their user experience represent practical stepping stones toward a more flexible grid, though fully integrated solutions such
as HEMS or local community management systems that coordinate multiple assets across households would deliver superior optimization outcomes.

Battery storage at the distribution level, spanning behind-the-meter home units, commercial installations, and grid-connected utility batteries, is one of the fastest-growing new load types in the Cerius-Radius network. Such storage is dual-natured: uncoordinated, it adds to peak demand as units charge during low-price hours that may coincide with local grid constraints, but coordinated against dynamic, location-aware signals it becomes a self-balancing resource that can flatten feeder loading and defer reinforcement. This dual character is precisely why the prioritization of connection requests is increasingly tied to a project's ability to balance itself locally: projects that can demonstrate self-balancing through on-site storage or controllable load can be moved ahead in the connection queue, since they impose smaller incremental burdens on constrained infrastructure. Denmark's revised connection-prioritization model, following 2025 EU Commission guidance, explicitly replaces strict first-come, first-served allocation with maturity, milestone, and ``grid-friendly'' criteria, the last of which rewards projects that relieve or at least do not worsen local capacity constraints \cite{Energinet2025Prioritering}. Dynamic tariffs are the signal that turns this latent capability into actual grid relief, ensuring that the rapid build-out of distribution-level batteries reduces rather than amplifies local congestion.

Together, these frameworks position Danish DSOs to deploy implicit flexibility mechanisms at scale. The regulatory question is no longer whether dynamic tariffs are permitted, but how best to design and implement them \cite{Madsen2024FlexibilityIncentivising}.

\section{Technical and Economic Justification}
\label{sec:technical}

Denmark's distribution grids face escalating pressures from widespread electrification and renewable energy integration. This section presents technical evidence of capacity constraints and quantifies the economic value of flexibility as an alternative to traditional reinforcement.

\subsection{Technical Justification: N-1 Constraints Under Pressure}

A fundamental requirement in distribution grid design is adherence to the N-1 security criterion, which mandates that networks must withstand the failure of any single major component without service interruption. Many grid components in the Cerius-Radius network currently operate near their thermal limits, reducing N-1 margins and increasing vulnerability to outages, accelerated component aging, and voltage instability.

It is worth noting that transformer capacity limits are not strictly binary thresholds. Dynamic Transformer Rating (DTR) methods allow operators to adjust the permissible load based on ambient conditions in real-time, oil temperature, and thermal history \cite{IEEE2012LoadingGuide,Bracale2020DTR,ariza2020a}. However, even with DTR, sustained overloading accelerates insulation degradation and reduces asset lifetime, making demand-side flexibility a valuable complement to dynamic rating approaches and proper use of batteries \cite{weckesser2021a}.

\begin{figure}[pos=!htb]
    \centering
    \includegraphics[width=0.80\linewidth]{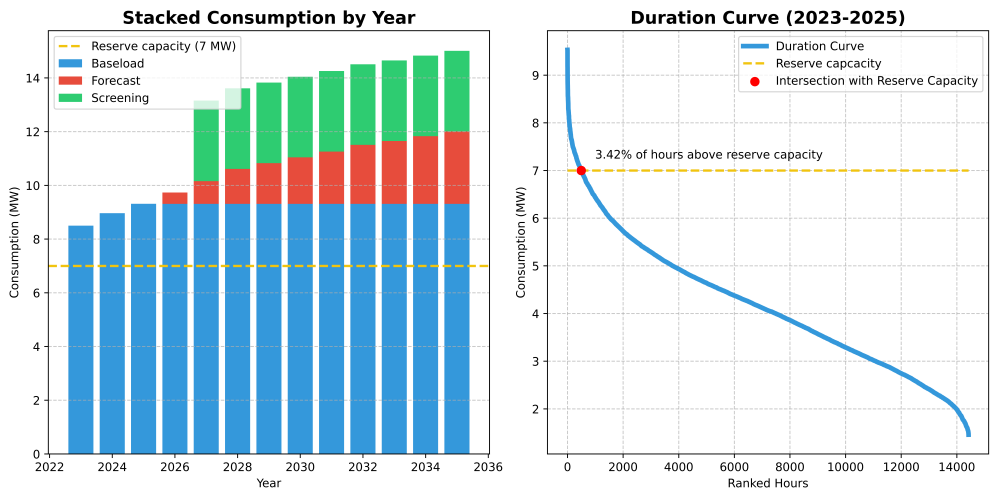}
    \caption{Demand development forecast (left) and sorted duration curve (right) for an anonymized primary substation. The 2027 screening reflects a 3.8~MW connection request that would push utilization beyond N-1 reserve margins for 3.42\% of annual hours.}
    \label{fig:DurationCurve}
\end{figure}

Figure~\ref{fig:DurationCurve} illustrates this challenge for a primary substation (50/10kV). The left panel shows forecast demand development through 2036, incorporating a screening request for connecting a 3.8 MW industrial facility in 2027. The right panel presents the corresponding sorted duration curve for 2023--2025, indicating that reserve capacity is insufficient for approximately 3.42\% of annual hours (approximately 300 hours per year).

The cost for a traditional reinforcement for this substation is estimated at 50 million DKK. The customer base primarily consists of residential users and small industries, precisely the segments where explicit flexibility procurement faces the greatest barriers and challenges (limited load magnitude, no organizational capacity for contracts, no facilities for verified dispatch).

Implicit flexibility mechanisms offer a viable alternative because they impose minimal requirements on customers and can be applied across diverse segments without requiring curtailment agreements. The 3.42\% of hours requiring flexibility is well within the response potential demonstrated by price-elastic customer segments.

This local picture should be read against the state of the national system, but the two must not be conflated. Denmark does not face a generation-adequacy problem: even during the coldest winter weeks, when the substation in Figure~\ref{fig:DurationCurve} approaches its limit, aggregate demand peaks at only around 6--7~GW and is met reliably through domestic generation and cross-border interconnection \cite{EnerginetSystemData2025}. The binding constraint is instead the \emph{network}, and it bites at every level: as the multi-year connection queues on the transmission system described in Section~\ref{sec:intro}, and, for the distribution operator, as individual feeders that breach their firm capacity. These network limits are local and time-specific in a way that energy scarcity is not. A single national energy price, such as the day-ahead spot price, signals only system-wide supply--demand balance and is largely blind to the congestion that actually binds at the substation. Dynamic, locally differentiated tariffs are precisely the instrument that exposes that constraint to the customers connected behind it, aligning their behaviour with the binding limit rather than the national average.

\subsection{Economic Benefits of Flexibility}

To evaluate the business case for flexibility-based approaches, we quantify the flexibility potential through a Net Present Value (NPV) comparison between traditional reinforcement and flexibility-based alternatives. Define:

\begin{equation}
NPV_{trad} = -\sum_{t=0}^{T} \frac{C^{trad}_{CAPEX,t} + C^{trad}_{OPEX,t}}{(1+r)^t}
\end{equation}

\begin{equation}
NPV_{flex} = -\sum_{t=0}^{T} \frac{C^{flex}_{CAPEX,t} + C^{flex}_{OPEX,t}}{(1+r)^t}
\end{equation}

where $C_{CAPEX,t}$ and $C_{OPEX,t}$ represent capital and operational expenditures in year $t$, $r = 5\%$ is the discount rate and $T = 10$ years is the investment horizon. The flexibility OPEX includes costs for tariff administration, customer communication, and a reserve fund to cover potential revenue shortfalls.

\subsubsection*{Case Study: Primary Substation Reinforcement Deferral}

We evaluate three scenarios for the 50 million DKK substation case.

\textbf{Scenario 1: Traditional Reinforcement (Baseline):} Immediate capital investment of 50 million DKK in year 0.
\[NPV_{trad} = -50.0 \text{ million DKK}\]

\textbf{Scenario 2: Flexibility Only:} No reinforcement; annual flexibility operational costs of 0.3 million DKK.
\[NPV_{flex} = -2.32 \text{ million DKK}\]
\[\Delta NPV = 47.68 \text{ million DKK}\]

\textbf{Scenario 3: Flexibility with Deferred Reinforcement:} Flexibility costs of 0.3 million DKK annually for years 1--10; reinforcement deferred to year 8.
\[NPV_{hybrid} = -36.16 \text{ million DKK}\]
\[\Delta NPV = 13.84 \text{ million DKK}\]

Table~\ref{tab:npv} summarizes the scenarios. Even the hybrid approach, which eventually requires full reinforcement, delivers substantial NPV benefits through deferral.

\begin{table}[htbp]
\centering
\caption{Comparative NPV Analysis of Investment Scenarios (50 million DKK Substation Case)}
\label{tab:npv}
\begin{tabular}{@{}lccp{4.5cm}@{}}
\toprule
Scenario & NPV (million DKK) & Net Benefit & Key Characteristics \\
\midrule
Traditional Reinforcement & $-50.00$ & Baseline & Immediate capital investment \\
Flexibility Only & $-2.32$ & 47.68 & Ongoing operational costs \\
Flexibility + Deferred & $-36.16$ & 13.84 & Reinforcement deferred 8 years \\
\bottomrule
\end{tabular}
\end{table}

Sensitivity analysis (Table~\ref{tab:sensitivity}) confirms robustness across discount rate and annual flexibility cost variations.

\begin{table}[htbp]
\centering
\caption{Sensitivity Analysis: Net Benefit vs. Traditional Reinforcement (million DKK)}
\label{tab:sensitivity}
\begin{tabular}{@{}lcccc@{}}
\toprule
Parameter & Low & Base & High & Net Benefit Range \\
\midrule
\multicolumn{5}{l}{\textit{Flexibility Only Scenario}} \\
Discount rate & 3\% & 5\% & 7\% & 47.4 -- 47.9 \\
Annual flex cost & 0.1 & 0.3 & 5.0 & 11.4 -- 49.2 \\
\midrule
\multicolumn{5}{l}{\textit{Deferred Reinforcement Scenario}} \\
Discount rate & 3\% & 5\% & 7\% & 8.0 -- 18.8 \\
Annual flex cost & 0.1 & 0.3 & 5.0 & 12.3 -- 15.4 \\
\bottomrule
\end{tabular}
\end{table}

\noindent\textit{Note: Annual flexibility costs are expressed in million DKK.}

\subsection{From Individual Response to Aggregate Reliability}

A common objection to implicit flexibility is that DSOs require operational certainty, while price signals cannot guarantee individual customer response. However, this objection misunderstands the statistical nature of aggregate demand response.

When flexibility comes from aggregated response across many customers, the law of large numbers applies. Individual customer response may be unpredictable, but aggregate response becomes statistically reliable as the customer base grows. Consider 1,000 customers with individual response probability of 60\%. The probability that fewer than 500 respond (50\% of potential) is less than 0.1\%. For DSOs managing large service territories, aggregate implicit flexibility can achieve high statistical reliability even with individually probabilistic response.

Furthermore, the assumption that price response requires active human decision-making is increasingly obsolete. Modern EVs, heat pumps, and smart appliances incorporate automated price response as a standard feature. Germany's \S14a EnWG now requires controllable loads such as wallboxes and heat pumps to be capable of receiving external control signals. Home Energy Management Systems (HEMS) translate price signals into automatic device adjustments without homeowner intervention.

The trajectory is clear: as the stock of price-responsive devices grows and automation becomes standard, the gap between "probabilistic" implicit flexibility and "guaranteed" explicit flexibility will narrow. DSOs investing in implicit flexibility infrastructure today are positioning for a future where automated response is the norm. The International Energy Agency's Net Zero Scenario requires a tenfold increase in demand response capacity by 2030 relative to 2020 levels, underscoring the central role that automated demand-side flexibility will play in future grid operations \cite{IEA2023DemandResponse}.

\subsection{Beyond Elasticity: Flexibility Functions for Dynamic Assessment}

The price elasticity analysis presented in Section~3 provides valuable insights into aggregate flexibility potential, but the concept of static elasticity has important limitations for operational grid management. Elasticity coefficients assume that customer response to price signals is constant over time, independent of prior consumption patterns, and deterministic in nature. In practice, flexibility is dynamic, state-dependent, and probabilistic.

Consider a supermarket with refrigeration systems: if the facility has already provided flexibility by pre-cooling during a low-price period, its ability to reduce consumption in subsequent hours is temporarily diminished until thermal mass is restored. Similarly, an electric vehicle that has provided flexibility by delaying charging may have reduced flexibility potential if the next trip is imminent. These dynamics cannot be captured by static elasticity measures.

\textbf{Flexibility Functions} offer a framework for addressing these limitations \cite{almassalkhi2026a,Madsen2023CharacterizingFlexibility}. A Flexibility Function characterizes the dynamic response of a flexible asset or aggregation of assets to varying price signals, capturing:

\begin{enumerate}
    \item \textbf{Temporal dynamics}: How flexibility potential evolves over time based on recent activation history and asset state
    \item \textbf{State dependencies}: How current flexibility depends on the state of underlying physical systems (temperature of the thermal mass of a building, battery state-of-charge, process buffers)
    \item \textbf{Probabilistic response}: Uncertainty quantification around expected demand response, enabling DSOs to assess reliability of aggregate flexibility
    \item \textbf{Constraints}: Physical and operational limits on flexibility provision (minimum run times, comfort bounds, process requirements)
\end{enumerate}

Drawing on the Flexibility Function framework, we can conceptually express the relationship between tariff signals and demand response as:

\begin{equation}
\mathcal{F}: \{\text{Tariff}(t)\}_{t=0}^{T} \times \mathbf{s}_0 \rightarrow P(D(t))
\end{equation}

where $\mathbf{s}_0$ represents the initial state of the flexible system and $P(D(t))$ is the probability distribution over demand realizations \cite{Junker2020StochasticFlexibility}.

This framework enables DSOs to move beyond deterministic assumptions about customer response. Instead of assuming that a given price signal will produce a fixed demand reduction, DSOs can obtain probabilistic assessments of likely demand trajectories, enabling more robust grid management decisions.

Flexibility Functions also provide a foundation for designing cost-reflective dynamic tariffs. By understanding how aggregate flexibility responds to price signals including the dynamic and state-dependent aspects, DSOs can design tariff structures that more effectively align customer behavior with grid constraints. Recent work published in the Danish Utility Regulator's anthology explores how Flexibility Functions can inform the design of distribution tariffs that achieve both cost-reflectiveness and operational effectiveness \cite{Madsen2024FlexibilityIncentivising}.

\subsection{Additional Value Streams}

The economic analysis above focuses on grid reinforcement deferral in the most directly quantifiable and operationally relevant value stream for DSOs. However, the full socio-economic case for demand-side flexibility extends beyond direct grid savings. Following DNV's framework for quantifying flexibility potential \cite{DNV2022FlexibilityEU} and recent work on incentivising multi-purpose flexibility \cite{Madsen2024FlexibilityIncentivising}, at least the following additional categories of avoided costs merit consideration:

\begin{itemize}
    \item \textbf{Avoided peaking generation costs}: Flexible demand can reduce reliance on expensive gas turbines during high-demand periods, delivering savings at the wholesale market level \cite{DNV2022FlexibilityEU}.
    \item \textbf{Avoided production losses from connection delays}: Grid constraints that prevent timely connection of new loads impose real economic costs. A data center flexibility study found that flexible connections enable 3--5 years faster access to power \cite{camus_datacenter}.
    \item \textbf{Reduced Value of Lost Load (VoLL) risk}: Sustained operation above rated capacity increases failure probability and accelerates equipment degradation. The Danish Energy Agency estimates VoLL at approximately 174 DKK per kWh of unserved energy \cite{Energistyrelsen2024VoLL}.
    \item \textbf{Renewable integration and curtailment reduction}: Flexible demand that absorbs surplus generation during periods of high renewable output reduces curtailment and supports decarbonization objectives. The same automated price response that enables congestion relief can simultaneously improve renewable utilization \cite{DNV2022FlexibilityEU,Madsen2024FlexibilityIncentivising}.
\end{itemize}

While these additional value streams are not quantified in detail here, they represent genuine societal benefits that further reinforce the investment case for flexibility. A comprehensive socio-economic assessment consistent with the ``flexibility-first'' evaluation approach mandated by EU Directive 2019/944 should account for these benefits alongside the direct grid savings demonstrated in our case study.

\section{Socio-Economic Considerations}

The deployment of DSF through dynamic tariffs raises equity and distributional considerations that must be addressed through complementary policy measures.

\subsection{Cost Allocation: Dynamic Tariffs vs. Traditional Approaches}

Cost-reflective tariffs shift network cost recovery toward customers who consume during peak periods or contribute to coincident demand. A concern raised in the literature is that such tariffs may disadvantage customers with limited ability to shift consumption, such as households without smart appliances or those with inflexible work schedules.
However, the alternative traditional reinforcement funded through uniform tariffs socializes infrastructure costs across all customers regardless of their contribution to peak demand. While Danish network charges already incorporate some time-of-use differentiation, the current structure does not reflect location-specific grid constraints. A customer who charges their EV during off-peak hours in an uncongested area and a customer who charges during peak congestion periods in a constrained area face similar network charges, despite imposing vastly different costs on the system.

Cost-reflective tariffs are not inherently less equitable than the status quo; they allocate costs differently. The equity question is whether targeted protections for vulnerable customers (discussed below) can address distributional concerns more effectively than the current cross-subsidization embedded in existing tariff structures.

\subsection{Access to Flexibility Benefits}

Customers with flexible assets, such as EVs, heat pumps, and battery storage, can benefit financially from dynamic tariffs by shifting consumption to low-price periods. Access to these assets correlates with income and housing type, potentially creating differential access to flexibility benefits.

This concern motivates complementary policy responses. First, \textbf{targeted support for vulnerable households} can ensure that efficiency gains from dynamic tariffs are not foregone entirely; instead, direct support mechanisms such as bill assistance programs or subsidized smart devices can help households that cannot participate in flexibility markets. Second, \textbf{consumer protection mechanisms} including tariff volatility caps, mandatory bill-impact forecasts for customers considering dynamic tariffs, and clear dispute resolution pathways can safeguard against unintended adverse effects. Such measures have proven effective in other jurisdictions implementing time-varying tariffs. 

\subsection{Investment Deferral: System-Wide Benefits}

Successful DSF deployment delivers three categories of system-wide benefits that extend beyond individual flexibility participants. First, deferred grid reinforcement investments translate directly into \textbf{lower electricity bills} for all customers, as avoided capital expenditure reduces the network charges that DSOs must recover through tariffs. The case study in this paper demonstrated potential savings of 13--48 million DKK per constrained substation, and the number of constrained substations is increasing substantially in these years of accelerating electrification.

Second, flexibility-based approaches enable \textbf{leaner grid planning}. Rather than sizing infrastructure for worst-case peak scenarios that occur during only a small fraction of annual hours, DSOs can plan for typical operating conditions while relying on flexibility to manage occasional peaks. This approach allocates capital more efficiently across the network.

Third, distributed flexibility resources enhance \textbf{system robustness}. A grid that can modulate demand in response to constraints is inherently more resilient to unexpected events, whether equipment failures, extreme weather, or rapid demand growth, than one relying solely on fixed infrastructure margins.

\section{Implementation Barriers, Challenges, and Policy Responses}

Several barriers and challenges must be addressed to achieve scalable deployment of implicit flexibility.

\subsection{Data Access and Privacy}

Granular metering data are essential for targeted tariff design and identification of flexible assets, yet GDPR mandates consent-based data sharing. Current opt-in frameworks result in low participation rates that hinder the development of the aggregator market.

\textbf{Policy response}: Shift toward ``data portability by default'' for certified market parties, preserving privacy while enabling aggregator participation. Aggregated forecasting at substation level, which avoids individual tracking, can support DSO congestion management without raising privacy concerns.

The previously mentioned concepts building on the Flexibility Function extended with automated grid-services enabled by automation, as described by the Smart-Energy OS principles in \cite{almassalkhi2026a}, implies that no real-time data are needed from the edge (e.g., household). In real-time the system operators need only the Flexibility Function for broadcasting tailor-made dynamic prices and tariffs. 

\subsection{Forecasting Uncertainty}

\textbf{Rebound effects}: Shifted loads may create secondary peaks if many customers respond similarly to price signals. Careful tariff design with gradual price transitions can mitigate this risk \cite{madsen2025a}.

\textbf{Non-stationarity}: Rapid electrification introduces consumption patterns not represented in historical data. Adaptive forecasting approaches incorporating real-time data updates and machine learning methods can address non-stationarity \cite{tohidi2025a}.

\textbf{From deterministic elasticity to probabilistic Flexibility Functions}: Traditional demand forecasting relies on deterministic elasticity estimates that assume constant customer response to price signals. As discussed in Section~5.4, this simplification becomes increasingly problematic as automation spreads and state-dependent effects become significant. Flexibility Functions provide a framework for probabilistic assessment of demand response, enabling DSOs to forecast not just expected demand but confidence intervals around that expectation. This probabilistic approach allows grid operators to maintain appropriate security margins while still capturing the benefits of aggregate flexibility. Implementing Flexibility Function-based forecasting requires investment in measurement infrastructure and modeling capabilities, but enables more sophisticated and reliable flexibility integration.

\subsection{Regulatory and Economic Barriers}

\textbf{CAPEX bias}: Traditional revenue-cap regulation allows DSOs to recover capital investments through their rate base while treating operational expenditures as costs to be minimized. This creates an implicit bias toward capital-intensive reinforcement over flexibility procurement, even when flexibility offers superior NPV.

Europe's continued CAPEX-driven regulatory approach still systematically favors infrastructure expansion over smarter operational system improvements. A shift to performance- and output-based regulation is essential for an efficient green transition at scale and to fully harness demand-side flexibility. 

\textbf{Policy response}: Adjust revenue-cap methodology to treat efficient flexibility procurement equivalently to capital investment (following the UK RIIO model's ``totex'' approach), removing implicit bias toward reinforcement.

\subsection{Policy Recommendations}

Based on the analysis presented in this article, we propose six policy measures to accelerate DSF deployment:

\begin{enumerate}
    \item \textbf{Flexibility-first mandate}: Require DSOs to document socio-economic assessment of non-wire alternatives before approving major reinforcement investments.
    
    \item \textbf{Baseline standardization}: Adopt uniform measurement and verification protocols for explicit flexibility markets to build confidence among consumers and aggregators.
    
    \item \textbf{Data democratization}: Introduce streamlined, GDPR-compliant data access pathways for pre-qualified aggregators, shifting from opt-in to secure opt-out frameworks.
    
    \item \textbf{OPEX neutrality}: Adjust revenue-cap methodology to treat efficient flexibility procurement equivalently to capital investment.
    
    \item \textbf{Consumer protection}: Implement tariff volatility caps, require bill-impact forecasts, and establish dispute resolution mechanisms to protect vulnerable households.

    \item \textbf{Progressive capacity pricing}: Introduce capacity-progressive tariff structures where consumption above defined thresholds incurs progressively higher network charges, strengthening incentives for peak demand reduction at the individual connection level.
\end{enumerate}

\section{Conclusion}

Denmark's distribution grids face capacity constraints that traditional infrastructure expansion cannot address quickly or cost-effectively enough. This article has argued that implicit demand-side flexibility through dynamic tariffs offers the necessary pathway for scalable, revenue-neutral solutions.
The argument rests on four empirically supported propositions:

First, \textbf{explicit flexibility mechanisms cannot scale to address system-wide congestion}. BNA (Begrænset Net Adgang - limited grid access) cannot relieve existing capacity constraints by definition; it operates only in buffer capacity. MAF (Market-based Activation of Flexibility) and similar contracted approaches face inherent transaction cost barriers that limit participation to large customers. For the residential and small commercial sectors where distributed flexible loads (EVs, heat pumps) are proliferating, implicit mechanisms represent the only viable and scalable approach.

Second, \textbf{price-responsive behavior already exists}. Industrial customers in the Cerius-Radius network demonstrate a strong inverse correlation between hourly prices and consumption ($\rho = -0.923$). Residential EV charging shows clear price responsiveness, even when that response creates local grid stress. The behavioral foundation for implicit flexibility is established; what is missing are \textit{price signals that reflect local grid constraints}.

Third, \textbf{the regulatory framework now enables deployment}. Market Model 3.0 and Tariff Model 3.0 explicitly support dynamic, cost-reflective distribution tariffs. Universal smart meter deployment provides the necessary data infrastructure. The question is no longer whether DSOs are permitted to implement dynamic tariffs but how to design them effectively.

Fourth, \textbf{flexibility achieves high reliability with modest requirements}. A data center flexibility study by Camus, encoord, and Princeton ZERO Lab demonstrates that grid power can remain available for more than 99\% of hours with on-site flexibility dispatched fewer than 70 hours annually \cite{camus_datacenter}. Combined with the automation trajectory, as electric vehicles, heat pumps, and smart appliances increasingly incorporate automated price response, the gap between probabilistic implicit flexibility and guaranteed explicit flexibility is rapidly narrowing.

The economic case is compelling: NPV analysis demonstrates potential savings of 13--48 million DKK per constrained substation compared to traditional reinforcement. The data center study found that flexible connections enable 3--5 years faster access to power while covering nearly 100\% of incremental system costs. These findings translate directly to distribution grid contexts.

Equity concerns are real, but can be addressed through targeted consumer protections rather than abandonment of efficiency gains. The alternative, traditional reinforcement funded through flat tariffs, socializes costs in ways that are also inequitable, while failing to provide the scalable flexibility that electrifying distribution grids require.

The path forward requires coordinated action: DSOs must develop tariff structures that accurately reflect local congestion; regulators must remove CAPEX bias in revenue-cap methodologies; policymakers must balance data access with privacy protection; and consumer protection mechanisms must ensure that flexibility benefits are broadly shared.

Implicit demand-side flexibility is not a complete solution. Explicit mechanisms retain important roles for large customers, emergency response, and TSO-level balancing services, where guaranteed dispatch remains essential. However, for the distributed and heterogeneous flexibility needs of electrifying distribution grids, implicit mechanisms through dynamic tariffs represent the necessary foundation. Denmark's regulatory framework has created the enabling conditions; the technical and economic evidence supports deployment. The remaining barriers are policy choices, not fundamental constraints.

\section*{Acknowledgments}
This work is supported by \textit{INSIEME} (Digital Europe No. 101194952),  \textit{ELEXIA} (Horizon Europe No. 101075656), \textit{ARV} (EU H2020 No. 101036723),  \textit{SEEDS} (Horizon Europe No. 10042024), \textit{IEA EBC - Annex 96 - Grid Integrated Control of Buildings} (EUDP Project No. 134251-549133) and the Green International Partnerships (GINP), Ministry of Higher Education and Science, Denmark (Ref. No. 2084-000478); The authors gratefully acknowledge the support.

\appendix

\section{NPV Calculations for the Cases}

\subsection{Calculations for Case 1 Scenario a and b}
\begin{figure}[pos=!htb]
    \centering
\includegraphics[width=0.85\linewidth]{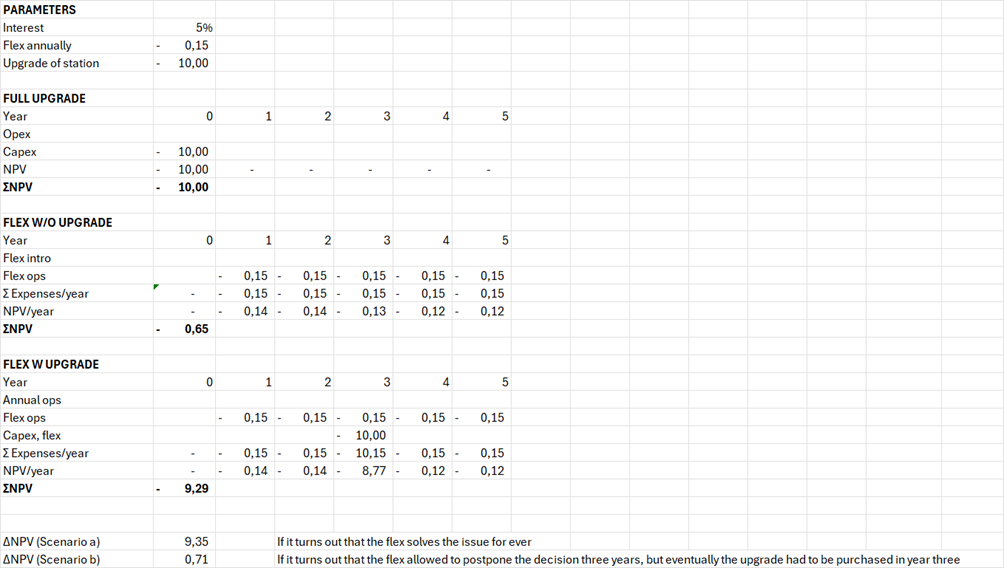}
\end{figure}

\newpage 

\subsection{Calculations for Case 2 Scenario a and b}
\begin{figure}[pos=!htb]
    \centering
\includegraphics[width=0.85\linewidth]{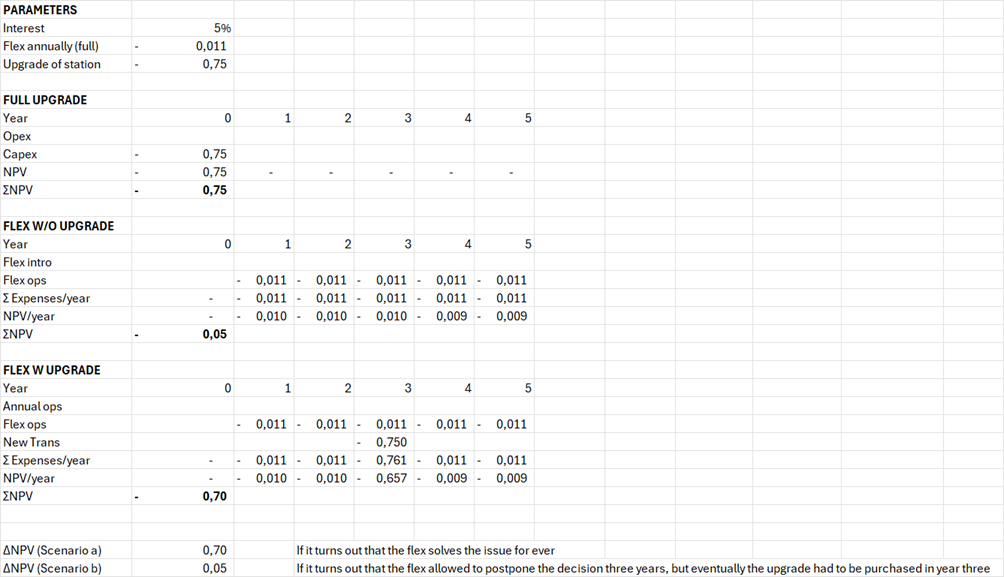}
\end{figure}

\subsection{Calculations for Case 3 Scenario a and b}
\begin{figure}[pos=!htb]
    \centering
\includegraphics[width=0.85\linewidth]{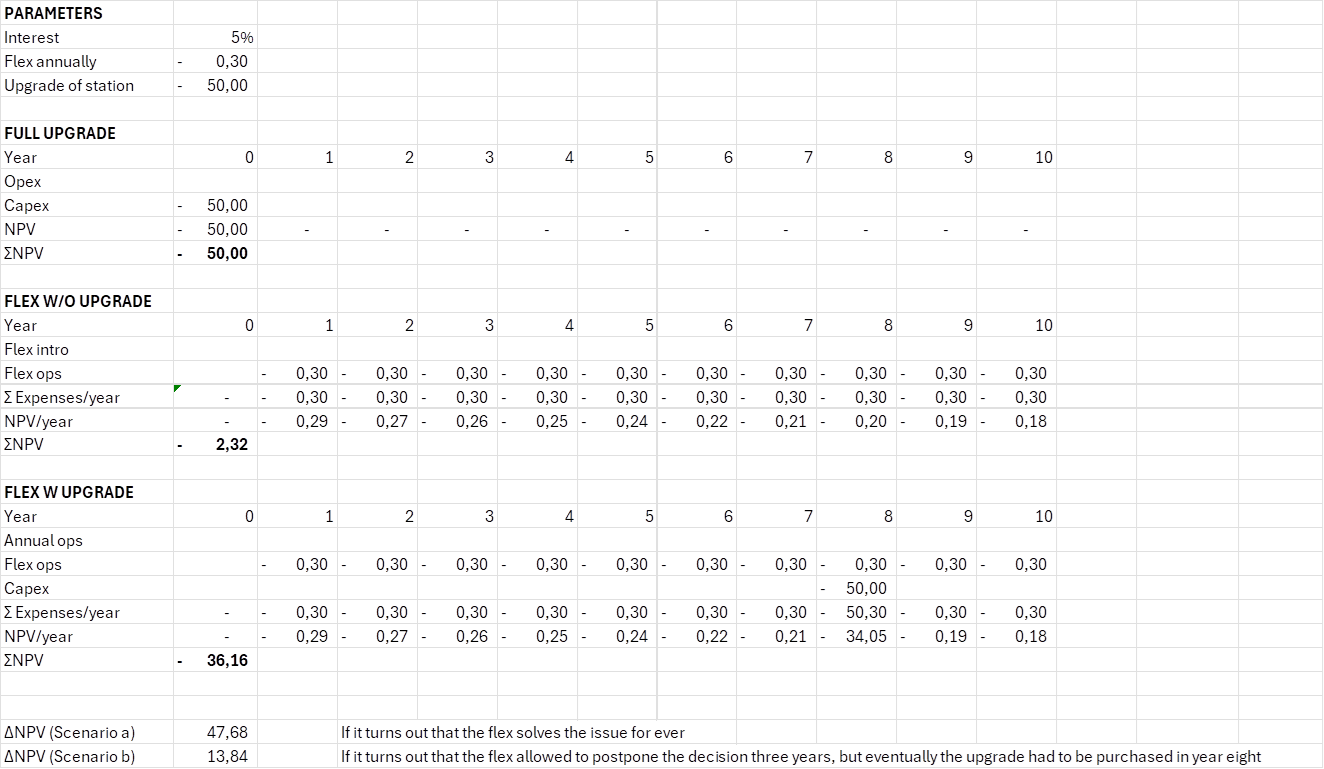}
\end{figure}

\printcredits

%% Loading bibliography style file 
%\bibliographystyle{model1-num-names}
%\bibliographystyle{cas-model2-names}
\bibliographystyle{elsarticle-num}

% Loading bibliography database
\bibliography{cas-refs}

%\vskip3pt

\section*{Bibliography}

\bio{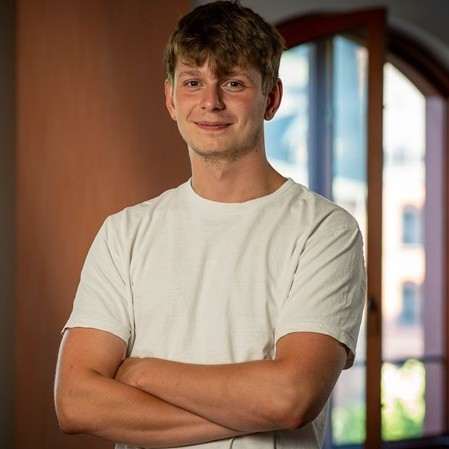}
Lucas Brylle is a research assistant at the Department of Applied Mathematics and Computer Science at the Technical University of Denmark (DTU), where his research focuses on demand-side flexibility and dynamic tariff design in Danish distribution grids. He holds a degree in Mathematical Modelling and Computation from DTU and works as a consultant at Deloitte on applied AI and data-driven product development. His interests lie at the intersection of energy market regulation, digitalization, and the green transition.
\endbio

\bio{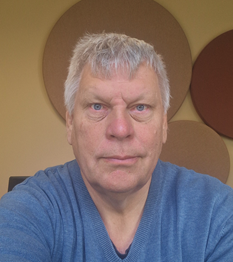}
Niels Andersen graduate as M.Sc. in Electrical Engineering from the Danish Technical University in 1988.  He was employed for 5 years as System Engineer in High Voltage DC transmissions at ABB. Later he joined the electrical utility SEAS-NVE and was involved in HVDC, grid studies and integrating of renewables at all voltage levels and consulting. Presently he is employed in the analysing department and do mainly analyses for grid development and integration of larger customers as well as power quality analysis at the utility company CERIUS-RADIUS.

\endbio

\bio{figs/pic3}
Henrik Madsen was appointed Professor in Mathematical Statistics at the Technical University of Denmark with a special focus on Stochastic Dynamical Systems in 1999. His main research interest is related to analysis, modelling and control of dynamical systems. The applications are mostly related to energy systems. He has authored or co-authored approximately 750 papers and 12 books. In 2016 he was appointed Knight of the Order of Dannebrog by Her Majesty the Queen of Denmark. In 2017 he was appointed Doctor HC at Lund University.
\endbio

\end{document}